\documentclass[10pt,conference]{IEEEtran}

\RequirePackage[T1]{fontenc}

\RequirePackage{graphicx}
\RequirePackage{mathptmx}
\RequirePackage{flushend}
\usepackage[utf8]{inputenc}
\usepackage{graphicx}
\usepackage{algorithm}
\usepackage{algpseudocode}
\usepackage{hyperref}

\usepackage{comment}

\usepackage{stmaryrd}
\usepackage{placeins}
\usepackage{balance}
\usepackage{tikz}
\usepackage{lmodern}
\usepackage{url}

\usepackage{newtxtext,newtxmath}

\usepackage{graphicx}
\usepackage{csquotes}
\usepackage{multicol}

\DeclareMathDelimiter{(}{\mathopen} {operators}{"28}{largesymbols}{"00}
\DeclareMathDelimiter{)}{\mathclose}{operators}{"29}{largesymbols}{"01}
\newcommand{\mathbbm}[1]{\text{\usefont{U}{bbm}{m}{n}#1}}
\newcommand{\mycomment}[1]{}

\algblock{Input}{EndInput}
\algnotext{EndInput}
\algblock{Output}{EndOutput}
\algnotext{EndOutput}

\usepackage{soul}

\AtBeginShipout{\AtBeginShipoutUpperLeft{
  \put(\dimexpr\paperwidth-1cm\relax,-1.5cm){\makebox[0pt][r]{\framebox{This paper was presented in the IEEE IPCCC conference, 2024, Orlando, FL}}}
}}

\begin{document}
\title{
Enhanced Outsourced and Secure Inference for Tall Sparse Decision Trees
}

\makeatletter
\newcommand{\linebreakand}{%
  \end{@IEEEauthorhalign}
  \hfill\mbox{}\par
  \mbox{}\hfill\begin{@IEEEauthorhalign}
}
\makeatother

\author{
\IEEEauthorblockN{Andrew Quijano}
    \IEEEauthorblockA{
    \textit{Dept. of Computer Science and Engineering} \\
    \textit{New York University}\\
    New York, NY, USA \\
    andrew.quijano@nyu.edu
    }
  \and
  \IEEEauthorblockN{Spyros T. Halkidis}
  \IEEEauthorblockA{
    \textit{Dept. of Applied Informatics} \\
    \textit{University of Macedonia}\\
    Thessaloniki, Greece \\
    halkidis@uom.edu.gr}
  \linebreakand
  \IEEEauthorblockN{Kevin Gallagher}
  \IEEEauthorblockA{
    \textit{NOVA LINCS} \\
    \textit{NOVA School of Science and Technology}\\
    Lisbon, Portugal \\
    k.gallagher@fct.unl.pt}
  \and 
  \IEEEauthorblockN{Kemal Akkaya}
  \IEEEauthorblockA{
    \textit{Advanced Wireless and Security Lab} \\
    \textit{Florida International University}\\
    Miami, FL, USA \\
    kakkaya@fiu.edu}
  \and
  \IEEEauthorblockN{Nikolaos Samaras}
  \IEEEauthorblockA{
    \textit{Dept. of Applied Informatics} \\
    \textit{University of Macedonia}\\
    Thessaloniki, Greece \\
    samaras@uom.edu.gr}
}

\maketitle          
%
\renewcommand{\thefootnote}{\fnsymbol{footnote}}
\footnotetext[1]{This paper was completed outside his job responsibilities at Amazon. The views expressed in this paper are those of the author and do not necessarily reflect the official policy or position of Amazon.}
\renewcommand{\thefootnote}{\arabic{footnote}} 

\begin{abstract}
A decision tree is an easy-to-understand tool that has been widely used for classification tasks. On the one hand, due to privacy concerns, there has been an urgent need to create privacy-preserving classifiers that conceal the user's input from the classifier. On the other hand, with the rise of cloud computing, data owners are keen to reduce risk by outsourcing their model, but want security guarantees that third parties cannot steal their decision tree model. To address these issues, Joye and Salehi introduced a theoretical protocol that efficiently evaluates decision trees while maintaining privacy by leveraging their comparison protocol that is resistant to timing attacks. However, their approach was not only inefficient but also prone to side-channel attacks. Therefore, in this paper, we propose a new decision tree inference protocol in which the model is shared and evaluated among multiple entities. We partition our decision tree model by each level to be stored in a new entity we refer to as a "level-site." Utilizing this approach, we were able to gain improved average run time for classifier evaluation for a non-complete tree, while also having strong mitigations against side-channel attacks.

\end{abstract}

%

\section{Introduction}
Machine learning is widely used in various real-world applications, such as managing student data \cite{student_data_dt}, analyzing health data \cite{PPDM}\cite{health_dt}, and fraud detection in energy consumption \cite{fraud_dt}. One type of classifier commonly used in machine learning is the Decision Tree (DT). A DT is a classifier that consists of decision nodes and leaves corresponding to an output class. DT evaluations have the advantage of being a fast and computationally inexpensive solution compared to other ML techniques. Also, a DT model is easy to audit as decisions are made through tree traversal.

However, as machine learning techniques begin to analyze sensitive data, such as medical records, privacy concerns arise. Privacy is becoming an important issue as there are regulations such as the Health Insurance Portability and Accountability Act (HIPAA) in the United States \cite{HIPAA} and the General Data Protection Regulation (GDPR) in the EU \cite{GDPR} that protect sensitive user data. One potential approach to balance both the desire to utilize machine learning and maintain a user's privacy is to use differential privacy. However, given that differential privacy relies on the injection of noise into the data set, the resulting classifier loses accuracy \cite{liuCloud}. Instead, we propose using a \textit{privacy-preserving decision tree (PPDT)} to achieve both accuracy and privacy preservation. A PPDT ensures that the server(s) storing the DT would not have information about the model leaked to the client, and the server is unaware of the client's input or output of the DT model. 

Previous work by Joye and Salehi \cite{JoyeAndSalehi} designed a theoretical PPDT classification protocol that used an encrypted integer comparison protocol based on Veugen's \cite{Veugen}\cite{Veugen_Correction} protocol that has additional protection against timing attacks. To this end, they converted a DT model into a complete binary tree, where all classes are in the bottom level of the tree. This ensures that every evaluation will have the same number of encrypted integer comparisons, so using a timing attack to extract information about the DT is still infeasible.

However, the PPDT model proposed in \cite{JoyeAndSalehi} has some issues. First, there is a performance loss since the full depth of the PPDT has to be traversed even if the classification could be computed earlier. In the case of a sparse DT \cite{sparse_tree}\cite{spare_tree_two}, which is defined as a DT that has few internal nodes, this implies that most leaves would not be at the maximum depth of the tree; therefore, Joye and Salehi's PPDT evaluation protocol would be inefficient. Also, they used an asymmetric based comparison based on RSA, which could be susceptible to power-based side channel attacks \cite{rsatiming}. Finally, since the classification occurs on one server, this server is a single point of failure. So, to address these concerns, we propose to split the PPDT model between multiple entities. We call these entities {\it{level-sites}}, inspired by the fact that the protocol progresses level by level, ending as soon as the client reaches a leaf. 

In this paper, we extend the theoretical PPDT classification protocol in \cite{JoyeAndSalehi} to a scenario where the DT model is shared between the level-sites, which improves our PPDT average run-time performance (especially for sparse, tall trees) while also providing robust security protections against side-channel attacks. 
To classify client data with our PPDT, a client would connect to the level-site $0$, and provide an encrypted feature vector to classify. Using \cite{JoyeAndSalehi} comparison protocol for numerical data or a timing attack resistant \cite{encrypted_equality}, the level-site $0$ calculates the index in scope for the next level in the PPDT and sends the index to the level-site $1$, etc. This process repeats until a level-site index is on a leaf and returns the classification to the client.

Assuming that the user's classification will not always result in traversing the whole tree, we observed a faster evaluation time in our experiments, as we have fewer encrypted comparisons to compute. Each level-site takes less than 1 second to complete the encrypted integer comparison and passing the index to the next level-site. We also show that our approach is resistant to side channel attacks and to single point of failures through our decentralized level-site approach.

\section{Related Work}
There have been several methods proposed for implementing PPDTs in the literature \cite{TruexEtAl}, \cite{BostEtAl}. Although some of these methods focus on protecting the confidentiality of training data \cite{LindellAndPinkas}, \cite{EmekciEtAl}, other related works, such as \cite{BostEtAl} and \cite{TaiEtAl}, focus on a secure inference protocol. A secure inference protocol enables users and model owners to interact so that the user obtains the prediction result while ensuring that neither party learns any other information about the user input or the model. 


Liu et al. \cite{liuCloud} designate Data Owners (DO) as the users who possess the DT training data, Cloud Service User (CSU) as a user with data to evaluate in the PPDT, and both Cloud Service Provider (CSP) and Evaluation Service Provider (ESP) as distinct cloud providers responsible for training and evaluating data, respectively. It is worth noting that both the CSP and the ESP must have knowledge of all the possible labels and attributes for the DT. They utilize the Distributed Two Trapdoors Public Key Cryptosystem (DT-PKC), which is similar to Paillier \cite{paillier1999public}, with the distinction that it allows comparisons between ciphertexts encrypted by different public keys. Liu et al. \cite{liuCloud} created new protocols to securely count the frequency of attributes in the data set, and modified Veugen's algorithm to be used for DT-PKC \cite{Veugen}\cite{Veugen_Correction}. Our paper has the advantage of using 2048-bit keys, which are more secure than Liu et al. \cite{liuCloud} and our PPDT has faster PPDT evaluation times.

Yuan et al. \cite{2023_ppdt_compare} utilize gradient-boosting decision trees (GBDT) to train DTs. They introduce the privacy-preserving distributed GBDT (PPD-GBDT), which uses differential privacy, polynomial approximation, and fully homomorphic encryption to achieve comprehensive privacy protection for their DT model. The DOs add noise to their local GBDT, and then convert the model into a polynomial format that is sent to the CSP. For evaluation, the client encrypts their input and sends the ciphertext and public key to the CSP. The CSP would return encrypted results, and the client would decrypt the classification. Our PPDT has generally faster PPDT evaluation times and no noise is added, which could potentially cause classification inaccuracies.

\section{Preliminaries}

\subsection{Background on Homomorphic Encryption}
An Additively Homomorphic Encryption scheme \cite{RivestEtAl} such as Paillier \cite{paillier1999public} or DGK \cite{DGK1}\cite{DGK2}\cite{DGK3} allows a user to combine two ciphertexts to receive a ciphertext output that has the sum of both input ciphertexts. For example, assume that there are two messages $\alpha$, $\alpha^{'}$ in the plaintext space ${\mathcal{M}}$. We write an encryption of $\alpha$ as $\llbracket \alpha \rrbracket$ and use the `boxplus' operator ($\boxplus$) to denote the addition of two ciphertexts. Then we have that, after decryption,
$\llbracket (\alpha + \alpha^{'}) \rrbracket$ = $\llbracket \alpha \rrbracket$ $\boxplus$ $\llbracket \alpha^{'} \rrbracket$. 

We use encryption to securely compare two encrypted integers. Assume that there are two integers held by A and B. Suppose that A has the $t$-bit integer $x=\sum_{i=0}^{t-1}x_i2^i$ in binary form, while party B has another $t$-bit integer $y=\sum_{i=0}^{t-1}y_i2^i$. The goal is for A and B, respectively, to obtain the protocol bits $\delta_A$ and $\delta_B$, such that $\delta_A \oplus \delta_B={\mathbbm{1}}\{x \le y\}$.

The secure integer comparison problem was first stated in \cite{Yao}, where two millionaires would like to find out who is richer without revealing the amount of their wealth. Various more advanced solutions followed in \cite{DGK1}\cite{DGK2}\cite{DGK3}. The high communication cost of these protocols was addressed by Veugen \cite{Veugen} and Joye and Salehi \cite{JoyeAndSalehi} improved Veugen's protocol to be resistant to timing attacks.

\subsection{System Model}
We assume the following entities.
A \textbf{Client} has a feature vector $\boldsymbol{x}$ and wants to classify it using the DT. The client creates public and private Paillier \cite{paillier1999public} and DGK \cite{DGK1}\cite{DGK2}\cite{DGK3} keys. The client will give the server the public keys. \textit{In a real-world scenario, a client could be a doctor who needs to know if a patient has thyroid disease}. A \textbf{Server} has access to the training data and creates the DT. The server creates the PPDT to send to the level-sites. To save network bandwidth, the server will give the level-sites the public keys. \textit{In a real-world scenario, this could be a laboratory that can inform a doctor if there is a probable sign of thyroid disease.} A \textbf{Level-site} only has a level of the PPDT and the public keys used to compare integers with the client. \textit{In a real-world scenario, a level-site would be a cloud provider hosting the PPDT}.

\subsection{Threat Model}
We assume the server is trusted, but the client and level-sites are ``honest-but-curious'' (HBC). The client may attempt to learn the decision nodes of a DT, which would contain the feature and a corresponding threshold. We assume that the level-sites will attempt to learn the client's feature vector data and classification output.

We expect the level-sites to strictly follow the protocol of our designed PPDT. However, we will assume that the level-sites would be curious to analyze both power consumption and computation times for potential side-channel attacks \cite{rsatiming}\cite{fhe_timing}.
We also consider a passive adversary, $\mathcal{A}$, outside the system in our model. $\mathcal{A}$'s goal is to learn the client input, the classification from the PPDT model, and the PPDT model's thresholds.

\section{PPDT Approach}

\subsection{Goals and Overview}
\label{sec:design_goals}
We are motivated by the fact that in a tall and sparse decision tree, most of the leaves in a DT are not at the maximum depth of the tree. Indeed, we computed the level of each classification from the training dataset \footnote{https://github.com/renatopp/arff-datasets} as shown in Table \ref{table:depth_analysis}. We want to minimize the encrypted comparisons used to improve our performance for our PPDT in this circumstance. In addition, if a level-site $l$ fails, we would like the traffic re-directed to a backup, solving the single point of failure in Joye and Salehi's approach \cite{JoyeAndSalehi}. 

\begin{table}[htb]
\vspace{-5mm}
\centering
\caption{Analysis of level of classifications of Training data}
\begin{tabular}{|l|l|l|l|l|l|}
\hline
Dataset         & Average   &   Median  & 3rd Quartile  &   Max & Size  \\ \hline
Hypothyroid     & 2.78      &   2       & 2             &   9   & 3372  \\ \hline
Spambase        & 9.60      &   9       & 11            &   22  & 4601 \\ \hline
Nursery         & 5.99      &   7       & 8             &   14  & 12960  \\ \hline
\end{tabular}
\label{table:depth_analysis}
\end{table}

\begin{figure*}[!htb]
\centering
\includegraphics[scale=1.0,height=4cm,width=\textwidth]{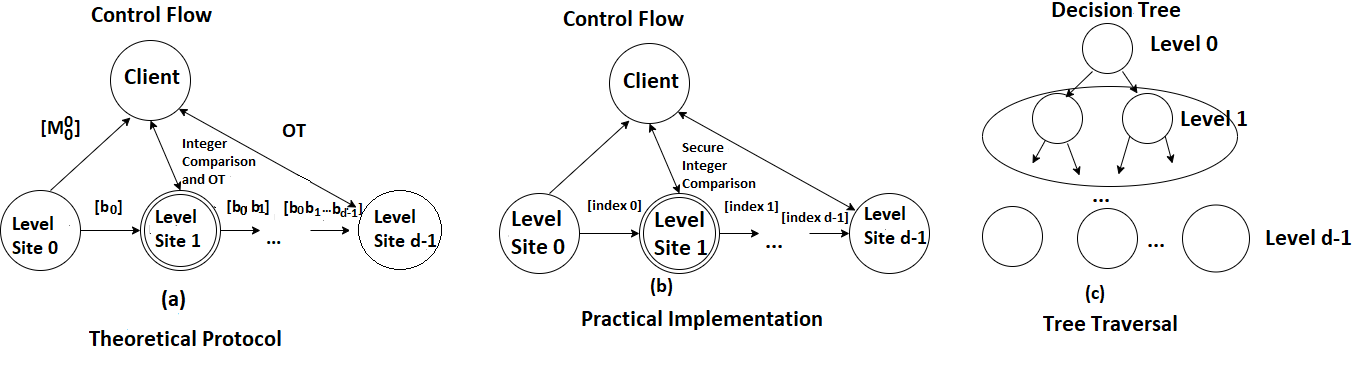}
\caption{
(a) Direct conversion of the Joye and Salehi PPDT \cite{JoyeAndSalehi} to the level-site approach.
(b) Control flow in our PPDT implementation, and
(c) Tree Traversal. We emphasize Level 1 of the DT evaluation}.
\label{fig:levels_in_control_flow}
\end{figure*}

Based on these goals, we set up our PPDT as follows: \textbf{(i)} The server builds the DT with the training data; \textbf{(ii)} The client sends to the server the DGK \cite{DGK1}\cite{DGK2}\cite{DGK3} and Paillier \cite{paillier1999public} public key. The server provides the client with both the classes of the DT; and \textbf{(iii)} For each level 0, 1, $\cdots$, $d - 1$, where $d$ is the depth of the DT, the server sends the \textbf{encrypted} thresholds and classifications to the respective level-site. 

\subsection{PPDT Inference}
In our setting, there is one client and $d$ level-sites, where $d$ is equivalent to the depth of the DT. The client has a private feature vector, $\boldsymbol{x}$ = ($x_1$, $x_2$, $\cdots$, $x_n$) $\in$ $\mathbb{Z}^n$, which is encrypted using homomorphic encryption to generate $\llbracket \boldsymbol{x}_i \rrbracket$, where $i \in {1, \cdots, n}$. We also assume that all thresholds are $t$-bits long. Each level site uses the Joye and Salehi encrypted integer comparison protocol \cite{JoyeAndSalehi} or a timing attack resistant encrypted equality protocol \cite{encrypted_equality} for categorical data. We note that in our PPDT, \textit{we improve our PPDT performance by not using oblivious transfer or converting to a complete binary tree as in the original work} \cite{JoyeAndSalehi}, as shown in Figure \ref{fig:levels_in_control_flow}.

\subsection{Level-site Evaluation}
We evaluated the PPDT by traversing down level-by-level, where level-site $l$ sends to level-site $l+1$ the encrypted index $i$ of the previous level-site. The client encrypts ${\boldsymbol{x}}$ and sends it to level-site 0.

Consider $l$ to be the level of the tree we are currently considering and $k$ to be the index of the current node at that level. We say $M_k^{l}$ is the blinded threshold comparison value at level l at position k and $\llbracket M_k^{l} \rrbracket$ to be the encrypted blinded threshold comparison value.

We have to compute at each level-site $l$ the quantity $\llbracket M_k^{l}\rrbracket = \llbracket x_{i} - T_{k}^{l} + \mu_l \rrbracket$. All these calculations are performed without decryption by using additively homomorphic encryption since: $\llbracket M_k^{l} \rrbracket = \llbracket x_{i} - T_k^{l} +\mu_l\rrbracket=$ $\llbracket x_{i} \rrbracket \boxplus \llbracket - T_k^{l} \rrbracket \boxplus \llbracket \mu_l \rrbracket$. We note that $\mu_l$ is a random $(t+\kappa)$ bit mask computed at the level-site $l$, where $\kappa$ is a security parameter \cite{JoyeAndSalehi}. In addition, $\llbracket x_{i} \rrbracket$ is the value of an attribute of ${\boldsymbol{x}}$ matched with the attribute of $T_k^l$.

Both the client and the level site $l$ utilize \cite{encrypted_equality} to compare categorical data or \cite{JoyeAndSalehi} to compare numerical data. We used a label encoder so that we can use an encrypted equality check to traverse a decision tree. The encrypted comparison protocols require $M_{k}^{l}$ and $\mu_l$ as inputs. At the end of the protocol, the level-site $l$ has $\delta_l$ and the client has $\delta_l^{'}$, such that: 

\[ \delta_l \oplus \delta_{l}^{'} = 
\begin{cases} 
        Numerical      & \mathbbm{1}\{ T_{k}^{l} \leq x_i \}\\
        Categorical    & \mathbbm{1}\{ T_{k}^{l} == x_i \} \\
\end{cases}
\]

Without the other corresponding bit, neither the level-site $l$ nor the client knows the value of the inequality. To get around this, the level-site $l$ can additively blind $\delta_l$ with $r^{'}$ and send it to the client to compute $(\delta_l$ $+$ $r^{'})$ $\oplus$ $\delta_l^{'}$. This output is returned to level-site $l$, as level-site $l$ knows $r^{'}$, it can determine the inequality output without revealing $\delta_l$ to the client. The output of the inequality informs how to update the index $k$ for the level-site $l + 1$. 

If the node at index $k$ at the level-site $l$ is a leaf, level-site $l$ returns the encrypted classification to the client. We summarize our approach in Algorithm \ref{alg:inference}.

\begin{algorithm}[htb]
\caption{PPDT Secure Inference Protocol with level-sites}
\label{alg:inference}
\begin{algorithmic}[1]
\State{next-index=0}
\State{Client sends $\mathbf{x}$ and next-index to level-site 0}
\For{\textit{int $l$ = 0; $l$ $<$ $d$; $l++$}}
    \State{level-site $l$ receives indexes and $\mathbf{x}$}
    \For{\textit{int $k$ = 0; $k$ $<$ $l.NodeCount()$; $k++$}}
        \If{\textit{Not in-scope based on next-index}}
            \State{\textbf{continue}}
        \ElsIf{in-scope and node $i$ is a leaf}
            \State{\Return{\textbf{encrypted-classification}} to client}
        \Else
            \State{secure comparison w/ client and level-site $l$}
            \State{next-index $\leftarrow$ set index for next level-site}
            \State{level-site $l$ sends index and $\mathbf{x}$ to level-site $l+1$}
            \State{\textbf{break}}
        \EndIf
    \EndFor
\EndFor
\end{algorithmic}
\end{algorithm}

\section{Security Analysis}
We construct security proofs via simulation, reducing our protocol to multi-party integer comparisons and homomorphic encryption, and citing original proofs. Assuming an honest-but-curious (HBC) adversary, we find that it can gather information on the PPDT through numerous queries. We then discuss countermeasures. We conclude by demonstrating how our PPDT is safeguarded against side-channel attacks and can be upgraded for post-quantum security.

Our proofs use simulation. We consider a PPDT classification protocol, $f(x,y)$, which we break into a two-party protocol such that $f(x,y) = f_{1}(x,y), f_{2}(x,y)$, and consider our implementation, $\pi$, of the functionality. During the execution of $\pi$, each of the two parties will generate a $view^{\pi}_{i}, \forall i \in \{1,2\}$. We claim our system is secure if we can show that there exist two simulators, $\mathcal{S}_{1}(1^{n}, x, f_{1}(x,y))$ and $\mathcal{S}_{2}(1^{n}, y, f_{2}(x.y))$ that are computationally equivalent to $view^{\pi}_{1}(x,y,n)$ and $view^{\pi}_{2}(x,y,n)$ where $n$ is a security parameter. 


\subsection{Security against an HBC client}
We assume a HBC client who follows the protocol faithfully but wishes to learn information about the tree node. To demonstrate that this client does not learn anything more than its view, we generate a simulator $\mathcal{S}_{1}$ that takes the input of the client and simulates the output of a level-site.


If the level-site $l$ is equal to zero, the simulator must sample a value $M_0^{0} \xleftarrow[]{R} \mathbb{Z}_p^{*}$, and encrypt it to achieve $\llbracket M_0^{0}\rrbracket$. This number is computationally indistinguishable from any real input that could be sent by a level-site given the properties of additive homomorphic encryption as demonstrated in \cite{RivestEtAl}. The client will then decrypt this number, resulting in $M_0^{0}$, and define $M_0^{\prime} = M_0^{0}$ and $m^{\prime}_0 = M_0^{0}$ mod $2^t$. To simulate the integer comparison protocol, it will then sample $b_r \xleftarrow[]{R} \{0,1\}$. This value will take the place of $b_0$\footnote{In our practical implementation, we use integers or indexes instead of bit strings that denote a path down the tree. However, this is functionally equivalent.}. As a result of the integer comparison protocol, each output in this case is equally likely to meet the standard of computational equivalence. 

In the case that the level-site is greater than 0, we need to perform more steps. First, the simulator sets $j \xleftarrow[]{}(b^{\prime}_0,\cdots,b^{\prime}_{l-1})_{2}$, according to the protocol. As before, the simulator samples a value $M_k^{l} \xleftarrow[]{R} \mathbb{Z}_p^{*}$ and follows the protocol by encrypting a randomly sampled value to receive $\llbracket M_k^{l}\rrbracket$, simulating the response from the level-site. The next level-site is then passed the encrypted client feature vector and the node index with which to compare. 
This process repeats until the simulator comes to a classification.

This simulator simply performs the same steps as the client, and randomly draws values to simulate the secret parameters of the level-site. Given that the comparison protocol and the additive homomorphic encryption scheme are statistically indistinguishable from the random output~\cite{Veugen,Veugen_Correction,encrypted_equality,RivestEtAl}, nothing about the internal state of the level-site is revealed through these operations. 


\subsection{Security against an HBC level-site l}
We assume the existence of HBC level-sites that wish to learn the client's private input vector $x$. We construct a simulator that produces output computationally indistinguishable from a legitimate run of the system.

The simulation begins when the simulator sets $x_i = 0, \forall x_i \in x$, generates a key pair $(pk, sk)$ and encrypts the $x_i$'s. The simulator then runs the protocol from the level-site by generating $\mu_l$, $\sigma_l$, $k$, $s$, etc., and generates $\llbracket M_k^{l}\rrbracket$. Then it generates $j$ by randomly selecting bits ${(b^{\prime}_0,\cdots,b^{\prime}_{l-1})}_2$. The simulator then goes through the comparison portion of the protocol. The level-site l passes on $\llbracket b_0,\cdots,b_l \rrbracket$ and the encrypted feature vector to level-site $l + 1$.

Simply, the simulator runs the level-site's protocol and encrypts the value 0 for $x_i, \forall x_i \in$ feature vector $\boldsymbol{x}$. It then steps through the protocol and, due to the encrypted integer comparison protocol, does not learn any information about the value of $x_i$. This makes it computationally indistinguishable from a real run of the system. Level-site $l$ uses the previous results of the tree-traversal for the previous level-sites, $0, \cdots, l-1$ the required $x_i$'s for the current level $l$ the thresholds for the current level $l$. Only the client knows the feature vector $\boldsymbol{x}$ and its classification. Thus, the level-site does not learn any relevant information from the client.

\subsection{Client Timing Attack and Performance Tradeoffs}
\label{sec:client_timing}
Over numerous runs, the client can begin to estimate the thresholds stored in different nodes by observing different feature vectors and classification results. An easy remediation is to throttle the client, as is done by other PPDT models \cite{liuCloud}. However, the client can determine the level on which the class is on using the amount of time that a classification takes. This attack can be mitigated by utilizing a proxy between the client and level-sites that adds random waiting times to the query result. 
The minimum extra time needed would be between 0.5 and 1 seconds to make it indistinguishable between two adjacent level-sites, as shown in Table \ref{table:performance_results}. Throttling would not severely affect the average run-time compared to the original Joye and Salehi PPDT protocol \cite{JoyeAndSalehi}.

\subsection{Security against Side Channel Attacks}
\label{sec:side_channel}

Side channel attacks are possible on multiple cryptosystems such as RSA~\cite{rsatiming}. We use Joye and Salehi's \cite{JoyeAndSalehi} comparison protocol for comparing numerical attributes, which is timing attack resistant on each level-site. For categorical data comparison, we used a modified \textit{Protocol 1 EQT-1} \cite{encrypted_equality} that always runs the same computations, which also makes it resistant to timing attacks. As both comparison protocols also have a similar run-time, the overall system has timing attack protection. Thus, the computation time between all the level-sites will be indistinguishable. If a classification finishes early, we can send a bogus value downstream to ensure that all level-sites complete a computation.

We use containerization to ensure stronger mitigations against power-based side-channel attacks. Containers are immutable after being created \cite{docker_immutable_and_resource_controls}, making it difficult to install software to analyze the runtime environment. Since containers are usually the end result of an automated deployment pipeline, consistent replacement of keys at level-sites \cite{key_switching_can_help} would also be a mitigation. More, a data owner could split the level-sites to be managed by multiple non-colluding cloud providers, which would make obtaining power readings to derive the private key much more difficult.


\section{Experimental Results}
\label{sec:results}
\subsection{Experiment Setup}
We implemented and tested the proposed approach using Amazon Elastic Kubernetes Service (EKS). We used t2.medium EC2 instances, which have 2 vCPUs and 4 GB of RAM. Our open source code implementation\footnote{https://github.com/adwise-fiu/Level-Site-PPDT} used the homomorphic encryption library implemented in Java \cite{AndrewQuijano} that implements Joye and Salehi's encrypted integer protocol \cite{JoyeAndSalehi} and Veugen's encrypted equality comparison \cite{encrypted_equality}.

We used Docker containers and Kubernetes as it allows the creation of replicas \cite{kubernetes_replicas} and supports horizontal scaling \cite{kubernetes_horizontal}, which fits our availability requirements in Section \ref{sec:design_goals}. We considered two EC2 VMs in the same region but in a different availability zone to simulate using two non-colluding cloud providers.

\subsection{Performance Results}
For our experiments, we ran our PPDT with each container storing a level-site. When obtaining our results, we did not implement throttling based on the discussion in Section \ref{sec:client_timing}, as we wanted to have a consistent view on how much time was needed for the PPDT evaluation at each level-site. We ran the evaluation 10 times each and provided the average execution time. The client experienced on average a 25-millisecond transmission delay for each connection, which is considered in the results. We also report the estimated time that Joye and Salehi PPDT would take by computing an evaluation that would traverse the $d$ levels of the PPDT.

Our solution agrees 100\% with a standard DT classification. Setting up the level-sites was completed in less than a second for all datasets. The performance results of our approach compared to Joye and Salehi are reported in Table \ref{table:performance_results}. Our superior performance is due to not requiring a complete binary tree \cite{JoyeAndSalehi}, reducing the evaluation depth and the number of encrypted comparisons.

\begin{table}[!htb]
\vspace{-5mm}
\caption{Comparing classification time of our approach with Joye and Salehi Approach under hypothyroid and nursery datasets.}
\centering
\begin{tabular}{|c|c|c|c|c|c|c|}
\hline
Levels & 
\begin{tabular}[c]{@{}l@{}}Hypothyroid\\ level-sites\end{tabular} & \begin{tabular}[c]{@{}l@{}}Hypothyroid\\ Joye \& Salehi \end{tabular} &
\begin{tabular}[c]{@{}l@{}}Nursery\\ level-sites\end{tabular} & \begin{tabular}[c]{@{}l@{}}Nursery\\ Joye \& Salehi \end{tabular} \\ \hline
2  & \footnotesize{\textbf{0.807 s}} & \footnotesize{3.279 s} 
   & \footnotesize{\textbf{0.656 s}} & \footnotesize{4.140 s} \\ \hline
   
4  & \footnotesize{\textbf{1.772 s}} & \footnotesize{3.279 s} 
   & \footnotesize{\textbf{2.041 s}} & \footnotesize{4.140 s} \\ \hline
   
9  & \footnotesize{4.148 s}          & \footnotesize{3.279 s} 
   & \footnotesize{\textbf{2.950 s}} & \footnotesize{4.140 s} \\ \hline
   
12 & \footnotesize{N/A} & \footnotesize{N/A} 
   & \footnotesize{5.648 s} & \footnotesize{4.140 s} \\ \hline
\end{tabular}
\label{table:performance_results}
\end{table}

Based on Table \ref{table:depth_analysis}, the average depth for the Hypothyroid set is 2.78 which corresponds to a value between 0.807 and 1.772 secs in Table \ref{table:performance_results}. This is significantly faster than the Joye and Salehi approach. Similarly, the average depth for Nursery data is 5.99 which corresponds to a value between 2.041 and 2.950 seconds, again much less than 4.140 of Joye and Salehi.

\subsection{Comparison with other Related Work}
We compared our approach with Li et al. \cite{liuCloud} and when repeating the same client queries it takes our PPDT 2.041, 6.198 and 15.878 seconds for the nursery, breast cancer and spambase dataset, respectively, which is at least 33\% faster. Compared to Yuan et al. \cite{2023_ppdt_compare} using the \textit{spambase} dataset, which they reported for evaluations. As shown in Table \ref{table:yuan_vs_level_sites}, our method is faster at almost all levels due to Yuan et al.'s need to encrypt data for each evaluation and the computational overhead of traversing the PPD-GBDT.


\begin{table}[!htb]
\vspace{-5mm}
\centering
\caption{Comparing the classification time of \cite{2023_ppdt_compare} with our approach using Spambase dataset.}
\begin{tabular}{|l|l|l|l|l|}
\hline
Levels      & Our Approach          & Yuan et al.\cite{2023_ppdt_compare}  \\ \hline
2           & \footnotesize{1.927 s}     & \footnotesize{0.49 s}  \\ \hline
3           & \footnotesize{2.736 s}     & \footnotesize{7.58 s} \\ \hline
4           & \footnotesize{3.937 s}     & \footnotesize{8.35 s} \\ \hline
5           & \footnotesize{4.638 s}     & \footnotesize{17.33 s}  \\ \hline
6           & \footnotesize{5.783 s}     & \footnotesize{19.37 s} \\ \hline
\end{tabular}
\label{table:yuan_vs_level_sites}
\vspace{-2mm}
\end{table}

\section{Conclusion}
In this paper, we present a new secure decision tree inference protocol by splitting the DT model into level-site constructs. Our PPDT exhibits optimal performance in sparse trees \cite{sparse_tree}\cite{spare_tree_two} due to fewer encrypted integer comparisons, resulting in faster performance compared to other PPDTs \cite{liuCloud}\cite{2023_ppdt_compare}\cite{JoyeAndSalehi}. In addition, it offers resistance to timing attacks through timing attack resistant comparison protocols \cite{JoyeAndSalehi}\cite{encrypted_equality}. Our PPDT also offers strong mitigations against power-based side channel attacks \cite{rsatiming}, as we use containers that offer power-based side channel protections \cite{docker_immutable_and_resource_controls}, key replacement \cite{key_switching_can_help}, and our approach can be used to split our PPDT with two non-colluding cloud providers. Finally, our PPDT protocols allow us to support load balancing to more evenly distribute more computing resources to higher levels of the PPDT, which would experience more usage and redundancy \cite{kubernetes_replicas} in case a level-site experiences an outage.

\section*{Acknowledgements}
We would like to thank Prof. Dimitrios Hristu-Varsakelis for helping at the initial stages of the paper by making suggestions. This work is supported by NOVA LINCS ref. UIDB/04516/2020 (\url{https://doi.org/10.54499/UIDB/04516/2020}) and ref. UIDP/04516/2020 (\url{https://doi.org/10.54499/UIDP/04516/2020}) with the financial support of FCT.IP.

\bibliographystyle{plain}
\bibliography{references}

\clearpage
\onecolumn
\thispagestyle{empty}  
\begin{center}
    \section*{IEEE Copyright Notice}
\end{center}
\noindent
{\normalsize
© 2024 IEEE. Personal use of this material is permitted. Permission from IEEE must be obtained for all other uses, in any current or future media, including reprinting/republishing this material for advertising or promotional purposes, creating new collective works, for resale or redistribution to servers or lists, or reuse of any copyrighted component of this work in other works.

Published in: \textit{Proceedings of the 2024 IEEE International Performance Computing and Communications Conference (IPCCC 2024)}.

DOI: \url{https://doi.org/10.1109/IPCCC59868.2024.10850192}
}
\clearpage
\twocolumn
\end{document}